\documentclass{article}
\usepackage{PRIMEarxiv}
\usepackage{graphics,graphicx}
\usepackage{amsmath}
\usepackage{amstext}
\usepackage{amssymb}
\usepackage{dcolumn}
\usepackage{amsfonts}
\usepackage{amsmath}
\usepackage{fixmath}
\usepackage{bm}
\usepackage{color} 
\usepackage{subfigure}
\newcommand{\beq}{\begin{eqnarray}}
\newcommand{\eeq}{\end{eqnarray}}
 \graphicspath{ {./FIG/} }
 \makeatletter
\def\blfootnote{\xdef\@thefnmark{}\@footnotetext}
\makeatother
\title{First-Principle Validation of Fourier's Law: One-Dimensional Classical Inertial 
Heisenberg Model}
\author{Henrique Santos Lima$^{a}$, Constantino Tsallis$^{b}$, and Fernando D. Nobre$^{c}$}
\begin{document}
\maketitle




\begin{abstract}

The thermal conductance of a one-dimensional classical inertial Heisenberg model 
of linear size $L$ is computed, considering the first and last particles 
in thermal contact with heat baths at higher and lower temperatures, 
$T_{h}$ and $T_{l}$ ($T_{h}>T_{l}$), respectively. These particles at extremities of 
the chain are subjected to standard Langevin dynamics, whereas all
remaining rotators ($i=2, \cdots , L-1$) interact by means of 
nearest-neighbor ferromagnetic couplings and evolve in time 
following their own equations of motion, being investigated numerically
through molecular-dynamics numerical simulations.  
Fourier's law for the heat flux is verified numerically with 
the thermal conductivity becoming independent of the lattice size 
in the limit $L \to \infty$, scaling with the temperature as 
$\kappa(T) \sim T^{-2.25}$, where $T=(T_{h}+T_{l})/2$.
Moreover, the thermal conductance, $\sigma(L,T)=\kappa(T)/L$, is well-fitted
by a function, typical of
nonextensive statistical mechanics, according to 
$\sigma(L,T)=A \exp_{q}(-B x^{\eta})$, where $A$ and $B$ 
are constants, $x=L^{0.475}T$, $q=2.28 \pm 0.04$, and $\eta=2.88 \pm 0.04$. 

\vskip \baselineskip


\end{abstract}   

\blfootnote{\textit{$^{a}$~Centro Brasileiro de Pesquisas Fisicas, Rua Xavier Sigaud 150, Rio de Janeiro-RJ 22290-180, Brazil.\\ 
E-mail: hslima94@cbpf.br }}

\blfootnote{\textit{$^{b}$~Centro Brasileiro de Pesquisas Fisicas and National Institute of Science and Technology of Complex Systems, Rua Xavier Sigaud 150, Rio de Janeiro-RJ 22290-180, Brazil \\
Santa Fe Institute, 1399 Hyde Park Road, Santa Fe, 
 New Mexico 87501, USA \\
Complexity Science Hub Vienna, Josefst\"adter Strasse 
 39, 1080 Vienna, Austria \\
E-mail: tsallis@cbpf.br}}

\blfootnote{\textit{$^{c}$~Centro Brasileiro de Pesquisas Fisicas and National Institute of Science and Technology of Complex Systems, Rua Xavier Sigaud 150, Rio de Janeiro-RJ 22290-180, Brazil 
E-mail: fdnobre@cbpf.br}}

\section{Introduction}

Two centuries ago, Fourier proposed the law for heat conduction
in a given macroscopic system, where the heat flux varies linearly with the gradient of 
temperature, $\mathbf{J}\propto -\nabla T$~\cite{Fourier1822}. 
For a simple one-dimensional system (e.g., a metallic bar along the $\hat{x}$ axis, 
${\mathbf{J}} = J{\bf{x}}$),
the heat flux $J$ (rate of heat per unit area) is given by

\vspace{-4mm}

\begin{equation}
J = - \kappa \, {dT \over dx}~, 
\label{one-dim-fourier-law}
\end{equation}

\noindent 
where $\kappa$ is known as thermal conductivity.  In principle, $\kappa$ may 
depend on the temperature, although most measurements are carried at room 
temperature, leading to values of $\kappa$ for many materials
(see, e.g., Ref.~\cite{cengelbook}). Usually, metals (like silver, copper, and gold)
present large values of $\kappa$, being considered as good heat conductors,
whereas poor heat conductors (like air and glass fibber) are characterized by 
small thermal conductivities; typically, the ratio between thermal
conductivities of these two limiting cases may differ by a $10^{4}$ 
factor. In most cases, good thermal conductors are also good
electrical conductors, and obey the Wiedemann-Franz law, which
states that the ratio of their thermal and electrical 
conductivities follow a simple formula, being directly proportional 
to the temperature~\cite{kittelbook}. 

In the latest years, many works were pursued for validating Fourier's law in 
a wide variety of physical systems, both experimentally and theoretically. 
Particularly, investigations for which microscopic ingredients may be responsible for
the property of heat conduction were carried, and it  
has been verified that thermal conductivity may be generated by different
types of particles (or quasi-particles). In the case of good electrical conductors
the most significant contribution to the thermal conductivity comes from free 
electrons, whereas in electrical insulators such contributions may arise 
from quasi-particles, like phonons and magnons, or even from defects.  
As examples, for antiferromagnetic electrical insulators such as $Sr_2$$CuO_3$ 
and $SrCuO_2$, magnons yield the most relevant contribution for the thermal 
conductivity, which can be fitted by a $1/T^2$ law, at high temperatures~\cite{Hlubek2012}. 
Several experimental investigations have verified Fourier's law 
in a large diversity of systems~\cite{Hlubek2012,Flumerfelt1969,
HuWenetal2015,Xu2016,Wu2017}, 
including coal and rocks from coalfields~\cite{HuWenetal2015}, as well 
as two-dimensional materials~\cite{Xu2016,Wu2017}.
On the other hand, some authors claim to have found anomalies~\cite{Hurtado2016}, 
or even violations of this law, for 
silicon nanowires~\cite{Yang2010}, carbon nanotubes~\cite{ZhiHan2011},
and low-dimensional nanoscale systems~\cite{LiuXu2012}.
Furthermore, a curious crossover, induced by disorder, was observed in quantum wires, where 
by gradually increasing disorder one goes from a low-disorder regime, where the law is 
apparently not valid,
to another regime characterized by a uniform temperature gradient inside the wire,
in agreement with Fourier's law~\cite{DubiVentra20091,DubiVentra20092}. 

From the theoretical point of view, many authors have investigated Fourier's law in 
a wide diversity of
models~\cite{Lebowitz1978,Buttner1984,Maddox1984,Wang1994,Laurencot1998,Aoki2000,%
Michel2003,Kawaguchi2005,Landi2014,Gruber2005,Bernardin2005,Bricmont20071,
Bricmont20072,%
Wu2008,Gaspard2008,Gers2010,Ezzat2011,DeMassi2011,Dahr2011,%
LiLiLi2015,OlivaresAnteneodo2016,LiLi2017,lima2023},
like a Lorentz gas~\cite{Lebowitz1978}, biological~\cite{Kawaguchi2005} and
small quantum systems~\cite{Michel2003}, chains of coupled harmonic~\cite{Landi2014}, 
or anharmonic~\cite{Aoki2000,Bricmont20071} oscillators, 
models characterized by 
long-range~\cite{Gers2010,OlivaresAnteneodo2016},
or disordered~\cite{Dahr2011} interactions, as well as systems 
of coupled classical rotators~\cite{LiLiLi2015,LiLi2017,lima2023}. 
In the case of a coupled XY nearest-neighbor-interacting rotator chain~\cite{LiLi2017}, 
the temperature dependence of the thermal conductance was well-fitted by a 
$q$-Gaussian distribution,

\vspace{-4mm}

\begin{equation}
\label{eq:qgaussian}
P_{q} (u) = P_{0} \exp_{q} (-\beta u^{2})~, 
\end{equation}

\noindent
defined in terms of the $q$-exponential function, 

\vspace{-4mm}

\begin{equation}
\label{eq:qexp}
\exp_{q} (u) = [1+(1-q)u]_{+}^{1/(1-q)}~; \quad  
\left( \exp_{1} (u) = \exp (u) \right)~,   
\end{equation}

\noindent
where $P_{0} \equiv P_{q}(0)$ and $[y]_{+}=y$, for $y>0$ (zero otherwise).  
The distribution in Eq.~(\ref{eq:qgaussian}) is very common in the context of 
nonextensive statistical mechanics~\cite{tsallisbook2023}, since it appears from
the extremization of the generalized entropy, 
known as $S_{q}$, characterized by a real index $q$~\cite{seminal_paper}, 

\vspace{-4mm}

\begin{equation}
\label{eq:sq}
S_{q}=k \sum_{i=1}^{W} p_i\left(\ln_q\frac{1}{p_i}\right)~, 
\end{equation}

\noindent
where we have introduced the $q$-logarithm definition, 

\vspace{-4mm}

\begin{equation}
\label{eq:qlog}
\ln_{q} u = \frac{u^{1-q} -1}{1-q}~; \qquad  
\left( \ln_{1} u = \ln u \right)~.  
\end{equation}

\noindent
Therefore, one recovers Boltzmann-Gibbs (BG) entropy,

\vspace{-4mm}

\begin{equation}
\label{eq:sbg}
S_{\rm BG}= - k \sum_{i=1}^{W} p_i \ln p_i~, 
\end{equation}

\noindent
as $\lim_{q \rightarrow 1} S_{q} = S_{\rm BG}$, 
whereas in the microcanonical ensemble, where all microstates present equal 
probability, $p_i=1/W$, Eq.~(\ref{eq:sq}) becomes,

\vspace{-4mm}

\begin{equation}
\label{eq:sqmicrocanonical}
 S_{q}=k\ln_{q}W~.  
\end{equation}

\noindent
Above, the $q$-exponential function in Eq.~(\ref{eq:qexp}) appears precisely as the 
inverse function of the $q$-logarithm of Eq.~(\ref{eq:qlog}), i.e.,  
$\exp_{q} (\ln_{q} u) = \ln_{q}(\exp_{q} (u)) = u$.

Since the introduction of the entropy $S_{q}$ in Eq.~(\ref{eq:sq}), a
large amount of works appeared in the literature defining generalized 
functions and distributions (see, e.g., Ref.~\cite{tsallisbook2023}).
Particularly, a recent study based on superstatistics, has found a 
stretched $q$-exponential probability distribution~\cite{maikepre2023},

\vspace{-4mm}

\begin{equation}
\label{eq:q-stretched-exponential}
P_{q} (u) = P_{0} \exp_{q} (-\beta |u|^{\eta})  \quad (0< \eta \leq 1), 
\end{equation}

\noindent 
as well as its associated entropic form.  

As already mentioned, the latest advances in experimental techniques made 
it possible to investigate
thermal and transport properties and consequently, Fourier's law, 
in low-dimensional (or even finite-size) systems, 
like two-dimensional materials~\cite{Xu2016,Wu2017},   
silicon nanowires~\cite{Yang2010}, carbon nanotubes~\cite{ZhiHan2011},
and low-dimensional nanoscale systems~\cite{LiuXu2012}.
These measurements motivate computational studies in finite-size systems of particles
that present their own equations of motion, e.g., systems of interacting 
classical rotators, whose dynamics may be followed through a direct 
integration of their equations of motion. In this way, one may validate (or not) 
Fourier's law, by computing the 
temperature and size dependence of the thermal conductance. 
A recent analysis of a system of coupled nearest-neighbor-interacting classical 
XY rotators~\cite{lima2023}, on $d$-dimensional lattices ($d=1,2,3$) of linear size $L$, 
has shown that, for a wider range of temperatures, the temperature dependence of 
the thermal conductance was better fitted by a more general Ansatz than the $q$-Gaussian
distribution of Eq.~(\ref{eq:qgaussian}). In fact, Fourier's law was validated in 
Ref.~\cite{lima2023} by fitting the thermal conductance in terms
of the functional form of Eq.~(\ref{eq:q-stretched-exponential}), 
with values of $\eta(d)>2$.

In the present work we analyze the thermal conductance of a one-dimensional 
classical inertial Heisenberg model of linear size $L$, 
considering the first and last particles in thermal contact with heat baths at 
temperatures $T_{h}$ and $T_{l}$ ($T_{h}>T_{l}$), respectively.
All remaining rotators ($i=2, \cdots , L-1$) interact by means of 
nearest-neighbor ferromagnetic couplings and evolve in time through
molecular-dynamics numerical simulations. 
Our numerical data validate Fourier's law, and similarly to those of Ref.~\cite{lima2023}, 
the thermal conductance is also well-fitted by the functional form of 
Eq.~(\ref{eq:q-stretched-exponential}). The present results suggest 
that this form should apply in general for the thermal conductance of 
nearest-neighbor-interacting systems of classical rotators. 
In the next section we define the model and the numerical procedure; in Section III we present
and discuss our results; in Section IV we pose our conclusions. 

\section{Model and Numerical Procedure}

The one-dimensional classical inertial Heisenberg model, for a system of $L$ interacting 
rotators, is defined by the Hamiltonian, 

\begin{eqnarray} 
\label{Heis}
\mathcal{H} = \frac{1}{2} \displaystyle{ \sum_{i=1}^{L} \bm{\ell}_i ^2+\frac{1}{2}
\sum_{\langle i j\rangle} \left(1-\mathbf{S}_i\cdot\mathbf{S}_j\right)}~, 
\label{Heisenberg}
\end{eqnarray} 

\noindent
where $\bm{\ell}_i \equiv ({\ell}_{ix},{\ell}_{iy},{\ell}_{iz})$ and 
$\mathbf{S}_i \equiv (S_{ix},S_{iy},S_{iz})$ represent, respectively, 
continuously varying angular momenta 
and spin variables at each site of the linear chain, 
whereas $\sum_{\langle i j\rangle}$ denote summations over pairs of nearest-neighbor spins; 
herein we set, without loss of generality, $k_{\rm B}$, moments of inertia, and ferromagnetic 
couplings, all equal to unit.
Moreover, spins present unit norm, $\mathbf{S}_{i}^{2}=1$, and at each site angular 
momentum $\bm{\ell}_i$ must be perpendicular to $\mathbf{S}_i$, yielding  
$\bm{\ell}_i \cdot \mathbf{S}_{i}=0$; these two constraints are imposed at the initial state and 
should be preserved throughout the whole time evolution. 

One should notice that, in contrast with a system of coupled
classical XY rotators, where canonical conjugate polar coordinates
are commonly used~\cite{lima2023}, in the Heisenberg case one often chooses
Cartesian coordinates~\cite{RapaportLandau1996,CirtoNobre2015,RodriguezCirtoTsallis2019}.
The reason for this is essentially technical, since 
in terms of spherical coordinates (more precisely, $\theta, \phi$ 
and their canonical conjugates $\ell_{\theta},\ell_{\phi}$), a troublesome term
$(1/\sin^2{\theta})$ appears in the corresponding equations of motion, 
leading to numerical difficulties~\cite{Evans77a,Evans77b}. 
However, some of the analytical results to be 
derived next recover those of the classical inertial XY model 
for $\mathbf{S}_i=(\sin{\theta_i}, \cos{\theta_i}, 0)$ and $\bm{\ell}_i=\ell_i\hat{\mathbf{z}}$.

It is important to mention that previous researches on the thermal conductivity 
have been carried either for a classical one-dimensional Heisenberg spin model,  
by using Monte Carlo and Langevin numerical simulations~\cite{Savin2005}, 
as well as for a classical one-dimensional spin-phonon system, 
through linear-response theory and the Green-Kubo formula~\cite{Savin2007}. 
These investigations did not take into account the kinetic contribution in Eq.~(\ref{Heis}), 
so that in order to obtain the thermal conductivity they assumed the validity of Fourier's law.
The main advantage of the introduction of the kinetic term in 
Eq.~(\ref{Heis}) concerns the possibility of deriving equations of motion, 
making it feasible to follow the time evolution of the system, through 
molecular-dynamics simulations, by a numerical integration of such equations. 
This technique allows one to 
validate Fourier's law, as well as to obtain its thermal conductivity directly.

\begin{figure}[h]
\includegraphics[width=1\textwidth]{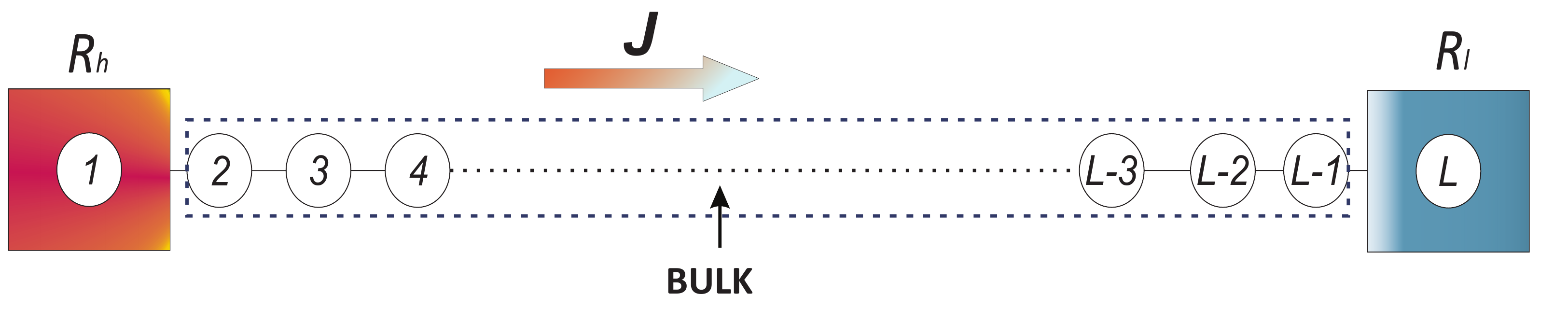}
\caption{Illustration of the system defined in Eq.~(\ref{Heis}), where the rotators at 
extremities of the chain are subjected to heat baths at different temperatures. 
The hot ($R_h$) and cold ($R_l$) 
reservoirs are at temperatures $T_{h}=T(1+\varepsilon)$ and $T_{l}=T(1-\varepsilon)$, respectively, 
leading to an average heat flux ${\mathbf{J}} = J{\bf{x}}$ throughout the bulk
(see text).
The rotators at sites $i=2, \dots, L-1$ interact with their respective nearest neighbors. 
}
\label{fig1}
\end{figure}

In order to carry on this procedure we consider an open chain of rotators with
the first and last particles 
in thermal contact with heat baths at higher and lower temperatures, 
$T_{h}$ and $T_{l}$ ($T_{h}>T_{l}$), respectively (cf. Fig.~\ref{fig1}), 
whereas all remaining rotators ($i=2, \cdots , L-1$) follow their usual 
equations of motion 
(see, e.g., Refs.~\cite{RapaportLandau1996,CirtoNobre2015,RodriguezCirtoTsallis2019}). 
In this way, one has for sites $i=2, \dots, L-1$, 

\vspace{-4mm}

\begin{align}
\begin{split}
\label{motion1a}
&\dot{\bf{S}}_i=\bm{\ell}_i \times \mathbf{S}_i~, \\
&\dot{\bm{\ell}}_i=\mathbf{S}_i \times (\mathbf{S}_{i+1}+\mathbf{S}_{i-1})~, 
 \end{split}
\end{align}

\noindent
whereas the rotators at extremities follow standard Langevin dynamics,   

\begin{align}
\begin{split}
\label{motion1b}
&\dot{\bm{\ell}}_1=-\gamma_h\bm{\ell}_1+\mathbf{S}_1 \times \mathbf{S}_2 + 
\mathbold{\eta}_h~, \\
&\dot{\bm{\ell}}_L=-\gamma_l\bm{\ell}_L+\mathbf{S}_L \times \mathbf{S}_{L-1} + 
\mathbold{\eta}_l~.\\
 \end{split}
\end{align}

\noindent
Above, $\gamma_h$ and $\gamma_l$ represent friction coefficients, whereas
$\mathbold{\eta}_{h}$ and $\mathbold{\eta}_l$ denote independent 
three-dimensional vectors, 
$\mathbold{\eta}_{h} \equiv (\eta_{hx},\eta_{hy},\eta_{hz})$, 
$\mathbold{\eta}_{l} \equiv (\eta_{lx},\eta_{ly},\eta_{lz})$, whose each Cartesian component
stand for a Gaussian white noise with zero mean and correlated in time, 

\begin{align}
\begin{split}
&\langle \eta_{h \mu}(t) \rangle = \langle \eta_{l \mu}(t) \rangle = 0~, \\
&\langle \eta_{h \mu}(t) \eta_{l \nu}(t') \rangle =   
\langle \eta_{h \mu}(t') \eta_{l \nu}(t) \rangle = 0~, \\
&\langle \eta_{h \mu}(t) \eta_{h \nu}(t') \rangle = 2 \delta_{\mu \nu} \gamma_{h}T_{h}
\delta(t-t')~, \\
&\langle \eta_{l \mu}(t) \eta_{l \nu}(t') \rangle = 2 \delta_{\mu \nu} \gamma_{l}T_{l}
\delta(t-t')~, \\
\end{split}
\end{align}

\noindent
with the indexes $\mu$ and $\nu$ denoting Cartesian components; from now on, 
we will set the friction coefficients $\gamma_h$ and $\gamma_l$ equal to unit. 

The condition of a constant norm for the spin variables yields 

\vspace{-4mm}

\begin{eqnarray} 
\label{norm-cond}
{d S_{i} \over dt} = {d \left( \mathbf{S}_i \cdot \mathbf{S}_i \right)^{1/2}  \over dt} = 0
\quad \Rightarrow \quad \mathbf{S}_i \cdot \dot{\bf{S}}_i = 0~, 
\end{eqnarray} 

\noindent
which should be used together with $\bm{\ell}_i \cdot \mathbf{S}_{i}=0$ in order to 
eliminate $\ddot{\bm{\ell}}_i$ and calculate $\ddot{\mathbf{S}}_i$ from 
Eqs.~(\ref{motion1a}) and (\ref{motion1b}).  
One has for rotators at sites $i=2, \cdots , L-1$,  

\vspace{-4mm}

\begin{align}
\begin{split}
\label{motion2a}
&\ddot{\mathbf{S}}_i = (\mathbf{S}_{i+1}+\mathbf{S}_{i-1}) -
\left[\mathbf{S}_i \cdot (\mathbf{S}_{i+1}+\mathbf{S}_{i-1}) + \dot{\bf{S}}_i^2 \right]\mathbf{S}_i~, 
\end{split}
\end{align}

\noindent
whereas for those at extremities,    

\vspace{-4mm}

\begin{align}
\begin{split}
\label{motion2b}
&\ddot{\mathbf{S}}_1= - \dot{\bf{S}}_1+ \mathbf{S}_2 -
\left[\mathbf{S}_1 \cdot \mathbf{S}_2 + \dot{\bf{S}}_1^2\right]
\mathbf{S}_1+ \mathbf{S}_1 \times \mathbold{\eta}_h~, \\
&\ddot{\mathbf{S}}_L= - \dot{\bf{S}}_L + 
\mathbf{S}_{L-1} - \left[\mathbf{S}_L \cdot \mathbf{S}_{L-1} +
\dot{\bf{S}}_L^2 \right] \mathbf{S}_L+\mathbf{S}_L\times \mathbold{\eta}_l~. 
 \end{split}
\end{align}

For the system illustrated in Fig.~\ref{fig1} we will consider the temperatures of the 
heat baths differing by $2\varepsilon$, with $\varepsilon$ representing a positive dimensionless 
parameter; moreover, 
the temperature parameter $T=(T_{h} + T_{l})/2$ will be varied in a certain range of 
positive values. The set of Eqs.~(\ref{motion2a}) and~(\ref{motion2b}) are transformed 
into first-order differential equations (e.g., by defining a new variable 
$\mathbf{V}_i\equiv \dot{\mathbf{S}}_i$) to be solved
numerically through Verlet's method~\cite{Verlet1967,PaterliniFerguson1998}, 
with a time step $dt=0.005$, for different lattice sizes $L$. 
The rotators at the bulk (i$=2, \cdots , L-1$) follow a continuity equation,

\vspace{-4mm}

\begin{eqnarray} 
\label{continuity-eq}
\frac{dE_i}{dt}=-(J_i-J_{i-1})~, 
\end{eqnarray} 

\noindent
where 

\vspace{-4mm}

\begin{eqnarray} 
\label{energy-i}
E_i = \frac{1}{2} \, \displaystyle{\bm{\ell}_i ^2+\frac{1}{2}
\sum_{j=i \pm 1} \left(1-\mathbf{S}_i\cdot\mathbf{S}_j\right)}~, 
\label{Heisenberg}
\end{eqnarray} 

\noindent
so the stationary state is attained for $(dE_i/dt)=0$, i.e.,  $J_i=J_{i-1}$.  

Data are obtained at stationary states, which, as usual, take longer
times to be reached for increasing lattice sizes. For numerical reasons, 
to decrease fluctuations in the bulk due to the noise, we 
compute an average heat flux by discarding a certain number of particles $p$ 
near the extremities (typically $p \simeq 0.15 L$). In this way, we 
define an average heat flux as 

\vspace{-4mm}

\begin{eqnarray} 
\label{av-heat-flux-a}
J \equiv {1 \over L-2p} \sum_{i=p+1}^{L-p}  \langle J_i \rangle~, 
\end{eqnarray} 

\noindent
with

\vspace{-4mm}

\begin{eqnarray} 
\label{av-heat-flux-b}
J_i = {1 \over 2} \left( \mathbf{S}_i \cdot \dot{\bf{S}}_{i+1}
   -\mathbf{S}_{i+1}\cdot \dot{\bf{S}}_i \right)~,  
\end{eqnarray}

\noindent
whereas $\langle .. \rangle$ denotes time and sample averages, to be described next.
Let us emphasize that for $\mathbf{S}_i=(\sin{\theta_i}, \cos{\theta_i}, 0)$ and 
$\bm{\ell}_i=\ell_i\hat{\mathbf{z}}$, one recovers the expression for the heat flux 
of the classical inertial XY model, i.e.,
$J_i=\frac{1}{2}(\ell_i+\ell_{i+1}){\sin{(\theta_i-\theta_{i+1})}}$~\cite{lima2023,mejia2019heat}, showing the appropriateness of the Cartesian-coordinate approach used herein for the 
classical inertial Heisenberg model. 

Let us now describe the time evolution procedure; 
for a time step $dt=0.005$, each unit of time corresponds to 
$200$ integrations of the equations of motion. 
We have considered a transient of $5 \times 10^{7}$ time units
to start computing the averages $\langle J_i \rangle$ in Eq.~(\ref{av-heat-flux-a}), 
checking that this transient time was sufficient to fulfill the condition
$J_i=J_{i-1}$ (within a three-decimal digits accuracy at least), 
for all values of $L$ analyzed.  
After that, simulations were carried for an additional interval of $2 \times 10^{8}$ 
time units (leading to a total time of $2.5 \times 10^{8}$ for each simulation). 
The interval $2 \times 10^{8}$ was divided into $80$ equally-spaced windows 
of $2.5 \times 10^{6}$ time units, so that time averages were taken inside each window;
then an additional sample average was taken over these $80$ time windows, 
leading to the averages $\langle J_i \rangle$.     

Using the results of Eq.~(\ref{av-heat-flux-a}) one may calculate the thermal 
conductivity of Eq.~(\ref{one-dim-fourier-law}), and consequently, the thermal
conductance,  

\vspace{-4mm}

\begin{eqnarray} 
\label{thermal-conduct}
\sigma=\frac{J}{T_h-T_l} = \frac{J}{2T \varepsilon} \equiv {\kappa \over L}~. 
\end{eqnarray} 

\noindent
In the next section we present results for both quantities, 
obtained from the numerical procedure described above.

\section{Results}

We simulated the system of Fig.~\ref{fig1} for different lattice sizes, 
namely, $L=50,70,100,140$, considering the heat-bath temperatures 
differing by $2\varepsilon$, with $\varepsilon=0.125$. The temperature 
parameter $T=(T_{h} + T_{l})/2$ was varied in the interval 
$0 < T \leq 3.5$, such as to capture both low- and high-temperature regimes. 
The values of $L$ ($L \geq 50$) were chosen adequately to guarantee that 
the thermal conductivity $\kappa$ did not present any dependence on the 
size $L$ in the high-temperature regime, as expected.

\begin{figure}
\centering
\subfigure[][]{\includegraphics[height=7.0cm]{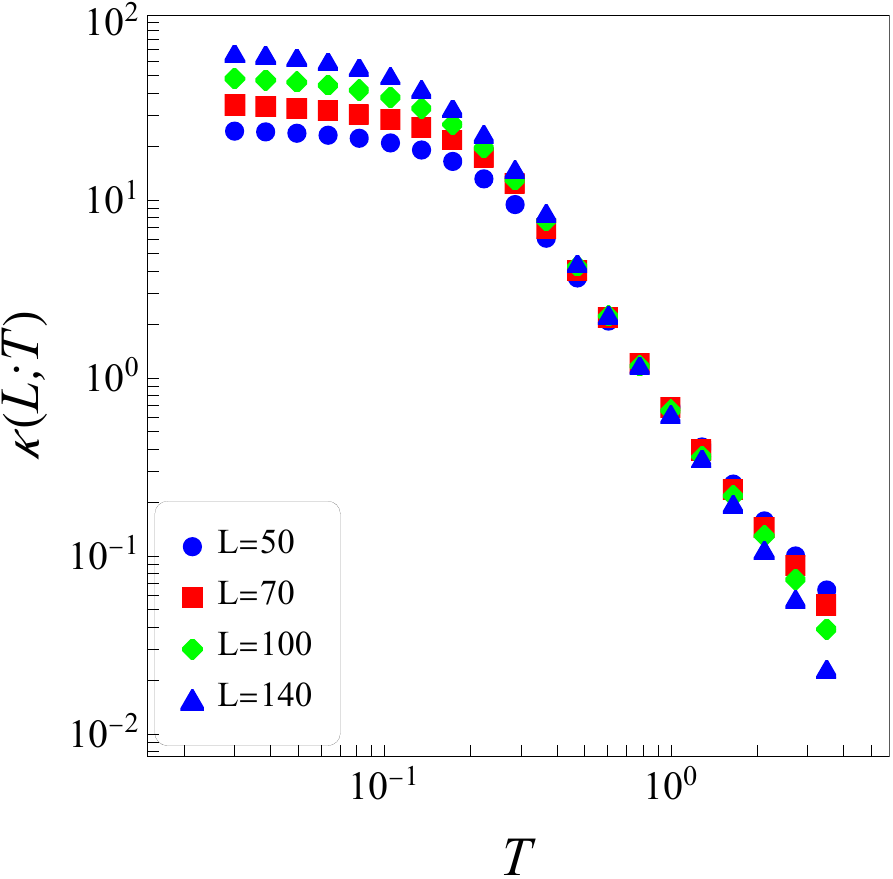}}
\hspace{0.8cm}
\subfigure[][]{\includegraphics[height=7.0cm]{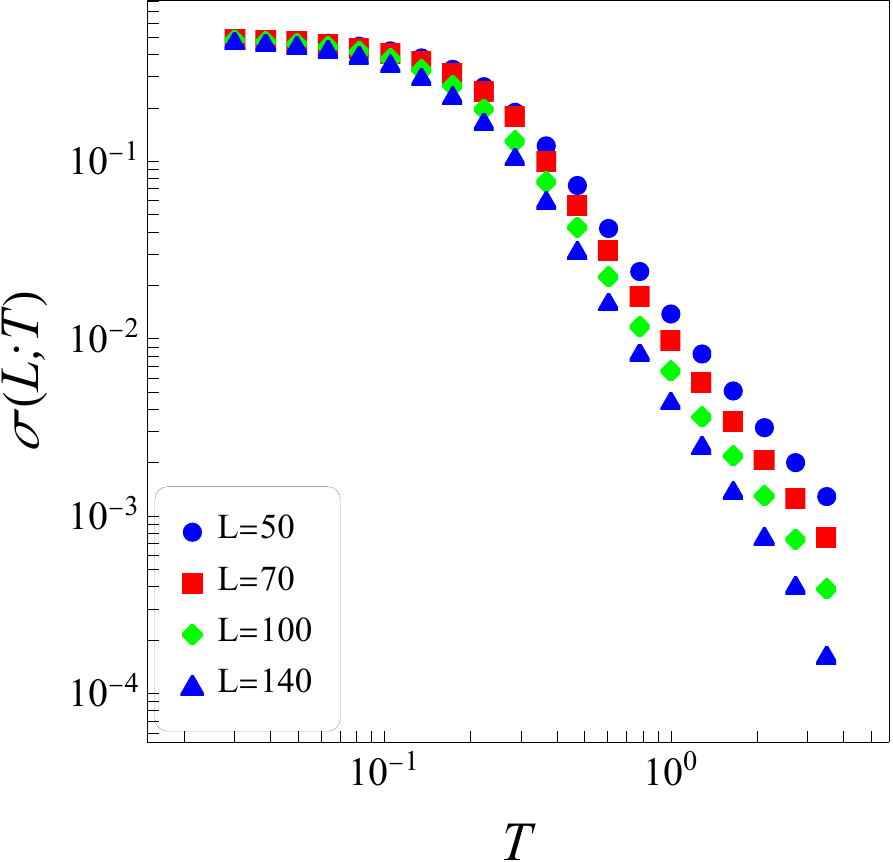}} \\
\caption{(Color online)
Numerical data for the thermal conductivity [panel (a)] and thermal conductance [panel (b)] 
are represented versus temperature (log-log plots) for different sizes ($L=50,70,100,140$) 
of the one-dimensional classical inertial Heisenberg model.   
One notices a crossover between the low- and high-temperature
regimes for $T \simeq 0.3$. 
As expected, higher temperatures amplify the effects of the Gaussian white 
noise, leading to larger fluctuations on numerical data, as shown clearly 
on panel (a). 
All quantities shown are dimensionless. 
}
\label{fig2}
\end{figure}

In Fig.~\ref{fig2} we present numerical data for the thermal conductivity [panel (a)]
and thermal conductance [panel (b)] versus temperature (log-log representations) 
and different sizes $L$. 
In Fig.~\ref{fig2}(a) we exhibit $\kappa(L,T)$ (the dependence 
of the thermal conductivity on the size $L$, used herein, will become clear below)
showing a crossover between two distinct regimes (for $T \simeq 0.3$), as described next.
(i) A low-temperature regime, where $\kappa$ depends on the size $L$,
decreasing smoothly for increasing temperatures ($L$ fixed).
The plots of Fig.~\ref{fig2}(a) show that, in the limit $T \to 0$, 
an extrapolated value, $\kappa(L,0) \equiv \lim_{T \to 0} \kappa(L,T)$, 
increases with $L$. This anomaly is attributed to the classical approach used herein, 
indicating that for low temperatures a quantum-mechanical procedure 
should be applied.
(ii) A high-temperature regime, where $\kappa$ essentially does not depend on 
$L$ (in the limit $L \to \infty$), as expected from Fourier's law. 
Moreover, in this regime one notices that $\kappa$ decreases with the
temperature as generally occurs with liquids and solids.  
For increasing temperature, the thermal conductivity of most liquids usually decreases 
as the liquid expands and the molecules move apart; in the case of solids, due to 
lattice distortions, higher temperatures make it more difficult for electrons to flow, 
leading to a reduction in their thermal conductivity. 
The results of Fig.~\ref{fig2}(a) indicate that 
the thermal conductivity becomes independent of the lattice size 
in the limit $L \to \infty$, scaling with the temperature as 
$\kappa(T) \sim T^{-2.25}$ at high temperatures.
In spite of the simplicity of the one-dimensional classical inertial Heisenberg model 
of Fig.~\ref{fig1}, the present results are very close to experimental verifications in 
some antiferromagnetic electrical insulators such as the Heisenberg chain
cuprates ${\rm Sr}_2{\rm CuO}_3$  
and ${\rm SrCuO}_2$, for which the thermal conductivity is well-fitted by a $1/T^2$ law 
at high temperatures~\cite{Hlubek2012}.

The same data of Fig.~\ref{fig2}(a) is exhibited in Fig.~\ref{fig2}(b) where 
we plot the thermal conductance $\sigma(L,T)=\kappa(L,T)/L$ versus temperature,
characterized by the two distinct temperature regimes described above. 
The low-temperature regime shows that the zero-temperature extrapolated 
value $\kappa(L,0)$ scales as $\kappa(L,0) \sim L$, leading to 
$\sigma(L,0) \equiv \lim_{T \to 0} \kappa(L,T)/L \simeq 0.5$.
On the other hand, in the high-temperature regime the 
thermal conductance presents  a dependence on $L$, as expected.

\begin{figure}
\centering
\includegraphics[width=0.7\textwidth]{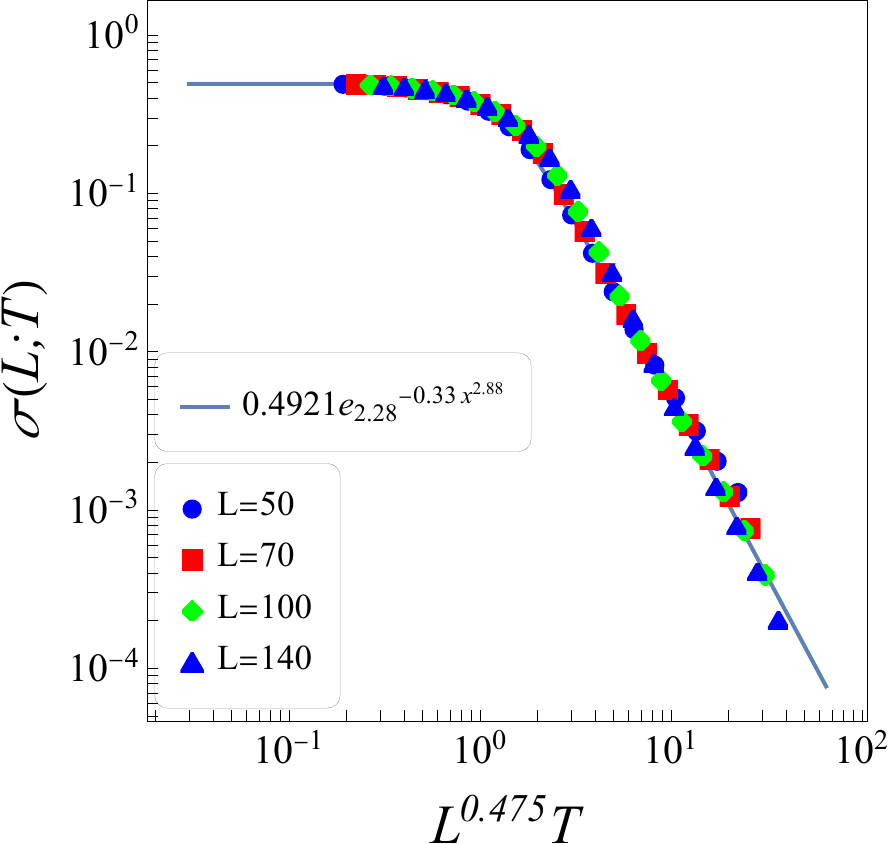}
\caption{
The plots for the thermal conductance of Fig.~\ref{fig2}(b) are shown in a log-log 
representation, for a conveniently chosen abscissa ($x=L^{0.475}T$), 
leading to a collapse of data for all
values of $L$ considered. The fitting (full line) is given by the function of  
Eq.~(\ref{eq:sigma-q-stretched-exponential}). 
}
\label{fig3}
\end{figure}

In Fig.~\ref{fig3} we exhibit the thermal-conductance data of Fig.~\ref{fig2}(b) 
in conveniently chosen variables, yielding a data collapse for all
values of $L$ considered. The full line represents essentially the form
of Eq.~(\ref{eq:q-stretched-exponential}), so that one writes

\vspace{-4mm}

\begin{equation}
\label{eq:sigma-q-stretched-exponential}
\sigma(L,T) = A \exp_{q}(-B x^{\eta})~, 
\end{equation}

\noindent
where $x=L^{0.475}T$, $q=2.28 \pm 0.04$, $\eta=2.88 \pm 0.04$, 
$A=0.492 \pm 0.002$, and $B=0.33 \pm 0.04$. 
Notice that this value of $\eta$ 
lies outside the range of what is commonly known as ``stretched'' 
[cf. Eq.~(\ref{eq:q-stretched-exponential})], so that the form above
should be considered rather as a ``shrinked'' $q$-exponential. 
It should be mentioned that in the case of coupled nearest-neighbor-interacting classical 
XY rotators on $d$-dimensional lattices ($d=1,2,3$)~\cite{lima2023},
the thermal conductance was also fitted by 
the form of Eq.~(\ref{eq:sigma-q-stretched-exponential}), 
with values of $\eta(d)>2$. Particularly, in the one-dimensional case,
such a fitting was attained for
$x=L^{0.3}T$, $q=1.7$, and $\eta=2.335$, showing that these numbers
present a dependence on the number of spin components ($n=2$, for XY
spins and $n=3$, for Heisenberg spins), as well as on the lattice
dimension $d$. 
By defining the abscissa variable of Fig.~\ref{fig3} in the general 
form $x=L^{\gamma(n,d)}T$, and using the $q$-exponential definition
of Eq.~(\ref{eq:qexp}), one obtains that the slope of high-temperature
part of the thermal-conductance data scales with $L$ as 

\vspace{-4mm}

\begin{equation}
\label{eq:sigma-L-scaling}
\sigma \sim L^{-[\eta(n,d) \gamma(n,d)]/[q(n,d) -1]}~, 
\end{equation}

\noindent
where we have introduced the dependence $(n,d)$ on all indices. 
Since the thermal conductivity ($\kappa=L \sigma$) should not depend
on the size $L$ (in the limit $L\to \infty$), Fourier's law becomes valid for 

\vspace{-4mm}

\begin{equation}
\label{eq:fourier-sigma-L-scaling}
{\eta(n,d) \gamma(n,d) \over q(n,d) -1} = 1~.  
\end{equation}

\noindent
The data of Fig.~\ref{fig3} lead to $[\eta(3,1) \gamma(3,1)]/[q(3,1) -1]=1.069 \pm 0.083$, 
whereas those for XY rotators on $d$-dimensional lattices yield
$1.0007, 0.95$,  and $0.93$, for $d=1,2$, and $3$, respectively~\cite{lima2023}, 
indicating the validation of Fourier's law for systems of 
coupled nearest-neighbor-interacting classical $n$-vector
rotators, through the thermal conductance 
form of Eq.~(\ref{eq:sigma-q-stretched-exponential}).

\section{Conclusions}

We have studied the heat flow along a one-dimensional classical inertial Heisenberg 
model of linear size $L$, by considering the first and last particles 
in thermal contact with heat baths at different temperatures, 
$T_{h}$ and $T_{l}$ ($T_{h}>T_{l}$), respectively.
These particles at extremities of 
the chain were subjected to standard Langevin dynamics, whereas all
remaining rotators ($i=2, \cdots , L-1$) interacted by means of 
nearest-neighbor ferromagnetic couplings and evolved in time following
their own classical equations of motion, being investigated numerically through 
molecular-dynamics numerical simulations.  

Fourier's law for the heat flux was verified numerically and 
both thermal conductivity $\kappa(T)$ and thermal conductance 
$\sigma(L,T)=\kappa(T)/L$ were computed, by defining 
$T=(T_{h}+T_{l})/2$. 
We have found that the slope of high-temperature
part of the thermal-conductance data scales with the system size as 
$\sigma \sim L^{-1.069}$, indicating that in the limit $L \to \infty$, one should get 
a thermal conductivity independent of $L$. Indeed, in this limit, we have found 
$\kappa(T) \sim T^{-2.25}$, for high temperatures. 
The whole thermal-conductance data was well-fitted
by the function $\sigma(L,T)=A \exp_{q}(-B x^{\eta})$, typical of
nonextensive statistical mechanics, where $A$ and $B$ 
are constants, $x=L^{0.475}T$, $q=2.28 \pm 0.04$, and $\eta=2.88 \pm 0.04$. 
Since the value of $\eta$ found herein lies
outside the range of what is commonly known as ``stretched'' 
$(0< \eta \leq 1)$, herein we called this fitting function of
a ``shrinked'' $q$-exponential. 
The present results reinforce those obtained recently for 
XY rotators on $d$-dimensional lattices~\cite{lima2023},
indicating that Fourier's law should be generally valid for systems of 
coupled nearest-neighbor-interacting classical $n$-vector
rotators, through the ``shrinked'' $q$-exponential function for 
the thermal conductance, with the indices $q(n,d)$ and  
$\eta(n,d)$ presenting a dependence on both number of
spin components and lattice dimension.

In spite of the simplicity of the model considered herein, the results for the 
thermal thermal conductivity at high temperatures ($\kappa(T) \sim T^{-2.25}$)
are very close to experimental verifications in 
some antiferromagnetic electrical insulators such as the Heisenberg chain
cuprates ${\rm Sr}_2{\rm CuO}_3$ and ${\rm SrCuO}_2$, for which the 
thermal conductivity is well-fitted by a $1/T^2$ law 
at high temperatures~\cite{Hlubek2012}.
Since nonextensive statistical mechanics has been used in the description 
of a wide variety of complex systems, one expects that the present results 
should be applicable to many of these systems in diverse 
non-equilibrium regimes.

\vskip\baselineskip
\noindent
{\bf Acknowledgements}

We have benefited from partial financial support by the Brazilian agencies 
CNPq and Faperj.


\end{document}